\documentstyle[sprocl]{article}

\def\mpl{\em Mod. Phys. Lett.} 
\def\np{\em Nucl. Phys.}
\def\pl{\em Phys. Lett.} 
\def\prl{\em Phys. Rev. Lett.}
\def\pr{\em Phys. Rev.}
\def\zp{\em Z. Phys.} 

\def\up{\uparrow}
\def\dw{\downarrow}

\def\be{\begin{equation}}
\def\ee{\end{equation}}
\def\bea{\begin{eqnarray}}
\def\eea{\end{eqnarray}}

\begin{document}

\title{SPIN AND ORBITAL ANGULAR MOMENTUM IN THE CHIRAL QUARK MODEL
\footnote{Talk given at 13th International Symposium on High Energy
Spin Physics, Protvino, Russia, September 8-12, 1998}
}

\author{X. SONG}

\address{Institute of Nuclear and Particle Physics, Department of
Physics\\
University of Virginia, Charlottesville, USA}


\maketitle
\abstracts{Quark spin and orbital angular momentum in the 
nucleon are calculated in symmetry-breaking chiral quark model. 
The results are compared with data and other models. }

{\bf Synopsis}: Recent measurements show (1) {\it strong flavor 
asymmetry of the antiquark sea} \cite{1}: $\bar d>\bar u$, 
(2) {\it nonzero strange quark content} \cite{2}: $<\bar ss>\neq 0$, 
(3) {\it sum of quark spins is small} \cite{3}: $<s_z>_{q+\bar
q}=0.1-0.2$, and (4) {\it antiquark sea is unpolarized} \cite{4}: 
$\Delta\bar u, \Delta\bar d\simeq 0$. All these features can be 
well and naturally understood in the chiral quark model ($\chi$QM) 
(for an incomplete list of references see [5,6]). 

{\bf The Chiral quark model}: The effective Lagrangian in the model 
is \cite{6}
$${\it L}_I=g_8{\bar q}\pmatrix{({\rm GB})_+^0
& {\pi}^+ & {\sqrt\epsilon}K^+\cr 
{\pi}^-& ({\rm GB})_-^0
& {\sqrt\epsilon}K^0\cr
{\sqrt\epsilon}K^-& {\sqrt\epsilon}{\bar K}^0
&({\rm GB})_s^0 
\cr }q, 
\eqno (1)$$
where $({\rm GB})_{\pm}^0=\pm {\pi^0}/{\sqrt 2}+{\sqrt{\epsilon_{\eta}}}
{\eta^0}/{\sqrt 6}+{\zeta'}{\eta'^0}/{\sqrt 3}$, $({\rm GB})_s^0=
-{\sqrt{\epsilon_{\eta}}}{\eta^0}/{\sqrt 6}+{\zeta'}{\eta'^0}/{\sqrt 3}$.
The breaking effects are explicitly included. $a\equiv|g_8|^2$ 
denotes the probability of chiral fluctuation or splitting 
$u(d)\to d(u)+\pi^{+(-)}$, and $\epsilon a$, the probability 
of $u(d)\to s+K^{-(0)}$ and so on. Since the coupling between the 
quarks and GBs is rather weak, the fluctuation $q\to q'+{\rm GB}$ 
can be treated as a small perturbation ($a\simeq 0.12-0.15$).

{\bf Spin, flavor and orbital contents}:
For a spin-up valence $u$-quark, the allowed fluctuations are
$$u_{\up}\to d_{\dw}+\pi^+,~~u_{\up}\to s_{\dw}+K^+,~~
u_{\up}\to u_{\dw}+({\rm GB})_+^0,~~u_{\up}\to u_{\up}.
\eqno (2)$$
The important feature of these fluctuations is that a quark {\it flips} 
its spin and changes (or maintains) its flavor by emitting a charged 
(or neutral) Goldstone boson. The light antiquark sea asymmetry 
$\bar d>\bar u$ is attributed to the existing {\it flavor asymmetry 
of the valence quark numbers} (two $u_v$ and one $d_v$) in the proton. 
The quark spin reduction is due to the {\it spin dilution} in the chiral 
splitting processes $q_{\up}\to q_{\dw}$+GB. Most importantly, the quark 
spin flips in the fluctuation with GB emission, hence the quark spin 
component changes one unit of angular momentum, $(s_z)_f-(s_z)_i=+1$ or
$-1$, the angular momentum conservation requires an {\it equal amount
change} of the orbital angular momentum (OAM) but with {\it opposite
sign}, i.e. $(L_z)_f-(L_z)_i=-1$ or $+1$. This {\it induced orbital 
motion} is distributed among the quarks and antiquarks, and compensates 
the spin reduction in the chiral splitting. Assuming that the valence
quark structure of the nucleon is SU(3)$_f\otimes$SU(2)$_s$, and
$\epsilon_\eta=\epsilon$, one obtains \cite{6}
$$\Delta u^p={4\over 5}\Delta_3-a,~~\Delta d^p=-{1\over 5}\Delta_3-a,~~
\Delta s^p=-\epsilon a,
\eqno (3)$$
The total spin carried by quarks and antiquarks is ${1\over 2}\Delta
\Sigma^p={1\over 2}(\Delta u^p+\Delta d^p+\Delta s^p)={1\over 2}
-a(1+\epsilon+f)$, where $f={1\over 2}+{{\epsilon}\over 6}
+{{\zeta'^2}\over 3}$. The best fit to the existing data (see Table 1) 
leads to $a\simeq 0.12-0.15$, $\epsilon\simeq 0.4-0.5$ and $\zeta'^2\simeq
0$, which gives $a(1+\epsilon+f)\simeq 0.25$. It means that about one half 
of the proton spin is carried by quarks and antiquarks. In addition, all
antiquark sea helicities are zero, $\Delta\bar q=0$ ($\bar q=\bar u,\bar d,
\bar s$) due to equal components of $\bar q_{\up}$ and $\bar q_{\dw}$ in
the GB. 

\begin{table}[ht]
\begin{center}
\caption{Quark spin and flavor observables in the proton in the
chiral quark model and nonrelativistic quark model (NQM).}
\begin{tabular}{cccc} \hline
Quantity & Data&  $\chi$QM  &NQM\\
\hline 
$\bar d-\bar u$ & $0.147\pm 0.039$ \cite{1},~$0.100\pm 0.018$ \cite{1} &
$0.130^*$ &0\\
${{\bar u}/{\bar d}}$ & $[{{\bar u(x)}\over {\bar d(x)}}]_{x=0.18}=0.51\pm
0.06$ \cite{1}, $0.67\pm 0.06$ \cite{1}& 0.68 &  $-$\\
${{2\bar s}/{(\bar u+\bar d)}}$ & 
$0.477\pm 0.051$ \cite{2}& 0.72 & $-$\\
${{2\bar s}/{(u+d)}}$ & 
$0.099\pm 0.009$ \cite{2} & 0.13 &0\\
 ${{\sum\bar q}/{\sum q}}$ & 
$0.245\pm 0.005$ \cite{2} & 0.24 & 0\\
 $f_s$ & $0.10\pm 0.06$ \cite{2}, $0.15\pm 0.03$ \cite{2} & 0.10 &  0\\
$f_3/f_8$ & $0.21\pm 0.05$ & 0.22 &1/3\\
\hline 
$\Delta u$ & $0.85\pm 0.05$ \cite{3} & 0.86 & 4/3\\
$\Delta d$&$-0.41\pm 0.05$ \cite{3} &$-$0.40&$-$1/3\\
$\Delta s$&$-0.07\pm 0.05$ \cite{3} &$-0.07$&0\\
$\Delta\bar u$, $\Delta\bar d$ & $-0.02\pm 0.11$ \cite{4} &0&0 \\
$\Delta_3/\Delta_8$ &2.17$\pm 0.10$&2.12& 5/3\\
$\Delta_3$&1.2601$\pm 0.0028$&1.26$^*$ &5/3\\
$\Delta_8$& 0.579$\pm 0.025$& 0.60$^*$ &1 \\
\hline
\end{tabular}
\end{center}
\end{table}

Using the hybrid quark-gluon model \cite{7} and assuming the induced
orbital motion is {\it equally shared} among quarks, antiquarks and 
gluon, one has
$$({\rm cos}^2\theta-{1\over 3}{\rm sin}^2\theta)[{1\over 2}-a(1-3k)
(1+\epsilon+f)]+<J_z>_{\rm G}={1\over 2}
\eqno (4)$$
where $\theta$ is the mixing angle and $k$ is the {\it sharing
factor}. If independent gluon degrees of
freedom are neglected, $<J_z>_{\rm G}=0$, $k=1/3$, $<L_z>_{q+\bar q}^p
=a(1+\epsilon+f)$, and $<J_z>_{q+\bar q}^p=1/2$. The missing part of 
the quark spin {\it is entirely transferred} to the orbital motion of 
quarks and antiquarks. If intrinsic gluon does exist, then $k<1/3$. 
The spin and OAM carried by quarks and antiquarks given in $\chi$QM
without the gluon and other nucleon models are listed in Table II. 
Extension to other octet and decuplet baryons were given in [6]. 
Including both spin and orbital contributions, the baryon magnetic 
moments are calculated. The result shows that the Franklin sum rule, 
$\mu_p-\mu_n+\mu_{\Sigma^-}-\mu_{\Sigma^+}+\mu_{\Xi^0}-\mu_{\Xi^-}=0$, 
still holds even the orbital contributions are included \cite{6}. 

{\bf Summary}: The $\chi$QM prediction on quark spin and flavor 
contents is in good agreement with the existing data. The 
probabilities of the chiral splittings $q\to q+\pi$, 
$q\to q+K(\eta)$, and $q\to q+\eta'$ are 10-15$\%$, 5-7$\%$ and
0.1$\%$ respectively. The OAM carried by quarks and antiquarks 
are given. The OAM contributions on the baryon magnetic moments
are also discussed. 

\begin{table}[ht]
\begin{center}
\caption{Quark spin and orbital angular momenta in different models.}
\begin{tabular}{cccccc} 
\hline 
 & NQM & MIT bag & $\chi$QM & CS \cite{8}&Skyrme\\ 
\hline 
$<s_z>^p_{q+\bar q}$ &1/2& 0.32 & 0.24 & 0.08&0\\
\hline 
$<L_z>^p_{q+\bar q}$ & 0 & 0.18 & 0.26 & 0.42& 1/2\\
\hline
\end{tabular}
\end{center}
\end{table}

\end{document}